\documentclass[a4paper,12pt]{article}%
\usepackage{amsfonts}
\usepackage{amsmath}
\usepackage{amssymb}
\begin{document}

\title{Non Abelian Vortices as Instantons on the Noncommutative Discrete space}
\author{\textsf{Hitoshi Ikemori } \thanks{ikemori@biwako.shiga-u.ac.jp}\\Faculty of Economics, Shiga University, \\Hikone, Shiga 522-8522, Japan
\and \textsf{Shinsaku Kitakado }\thanks{kitakado@ccmfs.meijo-u.ac.jp}\\Department of Physics, Faculty of Science and Technology,\\Meijo University,\\Tempaku, Nagoya 486-8502, Japan
\and \textsf{Hideharu Otsu } \thanks{otsu@vega.aichi-u.ac.jp}\\Faculty of Economics, Aichi University,\\Toyohashi, Aichi 441-8522, Japan
\and \textsf{Toshiro Sato} \thanks{tsato@mie-chukyo-u.ac.jp}\\Faculty of Law and Economics, Mie Chukyo University, \\Matsusaka, Mie 515-8511, Japan}
\date{}
\maketitle

\begin{abstract}
There seems to be close relationship between the moduli space of vortices and
the moduli space of instantons, which is not yet clearly understood from a
standpoint of the field theory. We clarify the reasons why many similarities
are found in the methods for constructing the moduli of instanton and vortex,
viewed in the light of the notion of the self-duality. We show that the
non-Abelian vortex is nothing but the instanton in $R^{2} \times Z_{2}$ from a
viewpoint of the noncommutative differential geometry and the gauge theory in
discrete space. The action for pure Yang-Mills theory in $R^{2} \times Z_{2}$
is equivalent to that for Yang-Mills-Higgs theory in $R^{2} $.

\end{abstract}

\newpage

\section{Introduction}

It is widely recognized that the exact solutions of field equations play an
important role in analyzing the properties of the field theory even in the
framework of a quantum theory. Getting solutions of the equation of motion in
a systematic way is important, especially in the case of non-Abelian gauge
theory or gauge coupled Higgs theory. Because existence of the effect of the
couplings even for the ground state is indispensable to understand the
significant properties of the theory such as symmetry breaking or confinement.
Although the classical solutions of the gauge theory were examined in various
models, the topological solitons are particularly interesting from a point of
view of the systematic construction of solutions.

The stability of such a solution is guaranteed by the topological properties
of soliton. The field equations of non-Abelian gauge theory or of gauge
coupled Higgs model are nonlinear second order differential equations, which
are not integrable in general. However, there exist the first order equations,
solutions of which automatically solve the second order field equations, and
these solutions have the properties of topological soliton. Such a topological
soliton equation is known as instanton equation for the Yang-Mills theory in
$R^{4}$ or BPS(Bogomol'nyi--Prasad--Sommerfield) equation in the case of
non-Abelian monopole in $R^{3}$\cite{BPS}.

The instanton equation for the Yang-Mills theory in 4 dimensional Euclidean
space $R^{4}$ is nothing but the self-duality equation for the field strength.
The BPS equation in 3 dimensional space describes the static non-Abelian
monopole of the Yang-Mills-Higgs theory, that is called BPS monopole, in the
limit of vanishing Higgs coupling. The BPS monopole equation can be derived as
a reduction of the self-dual Yang-Mills equation. Although other topological
solitons in the gauge theory are also known, those have a lot of common
properties. Generally, these solutions are called BPS solitons and the first
order equations, to which they obey, are called BPS equations. One of the
features of BPS equation which should be remarked is that we can minimize the
Euclidean action or the energy integral by completing the square with these
first order equations. The self-duality, although it changes its form in
various cases, is inherent as the common property.

The ADHM (Atiyah-Drinfeld-Hitchin-Manin) construction of instantons is one of
the most fruitful methods to obtain a soliton solution for the gauge
theory\cite{Atiyah:1978ri}. This method translates the instanton moduli space,
which is an information of the solutions of the self-dual Yang-Mills equation,
into the space of the solutions of the algebraic equation ( called hereafter
ADHM equation). The BPS monopoles are obtained by the similar method which was
proposed by Nahm\cite{Nahm:1979yw}. In this case, the monopole moduli are
determined by solving the first order ordinary differential equation (called
Nahm equation).

In the recent decade, we have had a glimpse of new aspect of the ADHM/Nahm
method. It was an interpretation as a configuration of D-branes. For example,
the composite system of $N$ D$4$-branes with $k$ D$0$-branes can be seen as a
configuration of $k$ instantons for $U(N)$ gauge theory in 4 dimensional space
identified with the bundle of the D$4$-branes. In this case, the self-dual
Yang-Mills equation and the ADHM equation are the conditions for supersymmetry
in D$4$-brane and D$0$-brane respectively. In the case of monopoles, the
system of $N$ D$3$-branes with $k$ D$1$-branes is interpreted as a
configuration of $k$ monopoles for $U(N)$ gauge theory in $3$ dimensions. The
BPS equation and the Nahm equation are the SUSY conditions in D$3$-brane and
D$1$-brane respectively. Here, in place of self-duality, a central role is
played by supersymmetry.

During the last half of this decade, there appeared a new family of solitons,
that is a non-Abelian vortex, adapted to the D-brane construction method of
moduli\cite{Tong}. The D-brane interpretation for the vortices can be given by
the configuration that consists of $N$ D$3$-branes suspended between two
parallel NS$5$-branes. As a result of study in this direction, it has been
pointed out that there seems to be close relationship between the moduli space
of vortices and the moduli space of instantons. The vortex moduli space, in
fact, involves half the elements of the ADHM construction and obeys the
relation similar to the ADHM condition. Although these results were surely
provided by a viewpoint of the D-brane and its supersymmetry, we do not
understand the vortex moduli and the relation to the ADHM from a viewpoint of
the field theory. The ADHM method allows us to construct the solutions to the
self-dual Yang-Mills equation in a systematic way. It is interesting to look
for a \textquotedblleft self-duality" in the case of the vortex described by
the \textquotedblleft half-ADHM"\cite{Eto}.

We cast some light on the notion of self-duality of the vortex to understand
the relation with the instanton. While the instanton equation expressed a
self-duality for the Hodge operator, the vortex equation seems to have no more
relation with the self-duality than that of being a first order BPS equation.
Actually, we understand that this equation does express a self-duality by
assuming appropriate space structure. We show that the non-Abelian vortex is
nothing but the instanton in $R^{2}\times Z_{2}$ space from the viewpoint of
noncommutative differential geometry and gauge theory in discrete
space\cite{Connes,Coquereaux,Morita,Varilly}. Such an idea has been once
proposed by Teo-Ting in the case of abelian model\cite{Teo:1997cn}. Here we
adapt this method to the case of non-Abelian vortex as an extension. Then we
clarified the reasons why many similarities are found in the methods for
constructing the moduli of instanton and vortex.

The constituents of this article are as follows. In section 2, we summarize
some properties of the vortex. In section 3, we explain differential geometry
and gauge theory in discrete noncommutative space. In section 4, we show the
fact that the non-Abelian vortices in $R^{2}$ can be considered as the
instantons in $R^{2}\times Z_{2}$ . Section 5 is assigned to the discussions.

\section{Non-Abelian vortex and self-dual BPS equation}

The vortex is a static solution of the Yang-Mills-Higgs system with a
translational symmetry in one direction \cite{Nielsen:1973cs}. The
configuration of the multi vortices consist of the individual elements with an
axial symmetry around itself. We can consider the vortex in the cross section
that is perpendicular to its axis. For example, we look upon the vortex in
$3+1$ dimensions as a model in $2+1$ dimensions. From this viewpoint, the
static vortex can be seen as a soliton solution in the 2 dimensional Euclidean space.

Let us summarize some properties of the vortex solution for Abelian Higgs
model \cite{Weinberg:1979er,Manton:1998kq,Schaposnik:2006xt}. The Lagrangian
of Abelian Higgs model in $2+1$ dimensions is given by
\begin{equation}
{\mathcal{L}}=\frac{1}{4}F^{\mu\nu}F_{\mu\nu}+\overline{D_{\mu}\phi}D^{\mu
}\phi+\frac{\lambda}{2}(|\phi|^{2}-c)^{2}\,.
\end{equation}
It is known that there are topologically stable static solutions in this model
in the case of $\lambda=1$. Such static solutions called vortices are the
configurations which minimize the energy integral%
\begin{equation}
E=\int_{R^{2}}d^{2}x\left(  \frac{1}{2}\left\vert F_{12}\right\vert
^{2}+\left\vert D_{1}\phi\right\vert ^{2}+\left\vert D_{2}\phi\right\vert
^{2}+\frac{1}{2}(|\phi|^{2}-c)^{2}\,\right)  \ ,
\end{equation}
by satisfying the BPS equations%
\begin{align}
iF_{12}\pm\left(  \left\vert \phi\right\vert ^{2}-c\right)   &  =0\ ,\\
\left(  D_{1}\pm iD_{2}\right)  \phi &  =0\ .\nonumber
\end{align}
The equations are often called vortex equations. We can also regard the energy
integral of the model in $2+1$ dimensional space-time as an action of the
Euclidean version of a theory in $1+1$ dimensions. In such a case, the
solutions of the BPS equations which give a minimum of the Euclidean action
are also called vortices in $2$ dimensional space.

In order to obtain the finite energy, it is necessary for these solutions to
satisfy the boundary conditions%
\begin{align}
|\phi|^{2}  &  \rightarrow c,\ D\phi\rightarrow0,\nonumber\\
F_{12}  &  \rightarrow0
\end{align}
at $\left\vert x\right\vert \rightarrow\infty$ the spacial infinity which is
identified with a circle $S^{1}$. This means that only pure gauge
configurations are allowed at the spacial infinity. Therefore, the global
properties of the solutions of the vortex equation are classified by the first
homotopy group%
\begin{equation}
\pi_{1}\left(  U\left(  1\right)  \right)  =\mathbb{Z\ }\text{,}%
\end{equation}
representing the topological mapping index for $S_{1}\rightarrow$ $U\left(
1\right)  .$ The integers corresponding to the elements of this homotopy group
are given by%
\begin{equation}
\dfrac{i}{2\pi}\int dx_{1}dx_{2}F_{12}=0,\pm1,\pm2,\cdots.
\end{equation}
This is nothing but the first Chern character of $U\left(  1\right)  $ gauge
field and is the topological charge of the solutions called vortex number.

The model can be extended to the Yang-Mills-Higgs model which has non-Abelian
gauge symmetry. Here we consider a $U\left(  N_{C}\right)  $ gauge group.
Generally, we can also extend the model to have a flavor symmetry among
$N_{F}$ Higgs fields. In this case, the energy integral is of the form
\begin{equation}
E=\int dx_{1}dx_{2}\mathrm{Tr}\left(  \frac{1}{2}\left\vert F_{12}\right\vert
^{2}+\left\vert D_{1}H\right\vert ^{2}+\left\vert D_{2}H\right\vert ^{2}%
+\frac{1}{2}\left(  HH^{\dag}-c\boldsymbol{1}_{N_{C}}\right)  ^{2}\right)
\end{equation}
provided by the Lagrangian in $2+1$ dimensions
\begin{equation}
{\mathcal{L}}=\mathrm{Tr}\left(  \frac{1}{4}F^{\mu\nu}F_{\mu\nu}+\left(
D_{\mu}H\right)  ^{\dag}D^{\mu}H+\frac{1}{2}(HH^{\dag}-c\boldsymbol{1}_{N_{C}%
})^{2}\,\right)  .
\end{equation}
Where, we define a covariant derivative $D_{\mu}$ and a field strength
$F_{\mu\nu}$ as
\begin{equation}
D_{\mu}=\partial_{\mu}+A_{\mu},\ F_{\mu\nu}=\partial_{\mu}A_{\nu}%
-\partial_{\nu}A_{\mu}+\left[  A_{\mu},A_{\nu}\right]  \ ,
\end{equation}
and $\mathrm{Tr}$ is a trace over the adjoint representation of $U\left(
N_{C}\right)  .$ It might be remarked that the gauge field $A_{\mu}$ and the
field strength $F_{\mu\nu}$ are $N_{C}\times N_{C}$ anti-hermitian matrices
and also that the Higgs field $H$ is represented by a $N_{C}\times N_{F}$
matrix which means an array of $N_{F}$ fundamental Higgs of $U\left(
N_{C}\right)  $.

The energy can be transformed into the form%
\begin{equation}
E=\int dx_{1}dx_{2}\mathrm{Tr}\left(  \frac{1}{2}\left\vert iF_{12}\pm\left(
HH^{\dag}-c\boldsymbol{1}_{N_{C}}\right)  \right\vert ^{2}+\left\vert \left(
D_{1}\pm iD_{2}\right)  H\right\vert ^{2}\right)  \pm i\int dx_{1}%
dx_{2}\mathrm{Tr}F_{12}\ ,\
\end{equation}
omitting a surface integral which has no affect on account of the boundary
conditions. The BPS equations minimizing the energy are
\begin{align}
iF_{12}\pm\left(  HH^{\dag}-c\boldsymbol{1}_{N_{C}}\right)   &
=0\ ,\nonumber\\
\left(  D_{1}\pm iD_{2}\right)  H  &  =0\ ,
\end{align}
in this case. These equations also have topologically stable solutions in a
similar way as in the Abelian case, which we call non-Abelian
vortices\cite{Tong,Eto}. It is also obvious that pure gauge configurations are
allowed at the spacial infinity $\left\vert x\right\vert \rightarrow\infty$ .
It means that the topological property of the non-Abelian vortices is
classified by the mapping index for $S_{1}\rightarrow$ $U\left(  N_{C}\right)
.$ On account of the fact that $U\left(  N_{C}\right)  \ $is equal to
$U\left(  1\right)  \times SU\left(  N_{C}\right)  $, the corresponding
homotopy group is
\begin{equation}
\pi_{1}\left(  U\left(  N_{C}\right)  \right)  =\pi_{1}\left(  U\left(
1\right)  \right)  =\mathbb{Z}%
\end{equation}
whose elements are integers and are identified as vortex numbers given by%
\begin{equation}
\dfrac{i}{2\pi}\int dx_{1}dx_{2}\mathrm{Tr}F_{12}=0,\pm1,\pm2,\cdots\ .
\end{equation}

Although this model has the local $U\left(  N_{C}\right)  $ gauge symmetry and
the global $SU\left(  N_{F}\right)  $ flavor symmetry, there occurs the
symmetry breaking due to the existence of the Higgs potential. The vacuum of
the theory has completely broken symmetries and there appear vortex solutions,
provided that $N_{F}\geq N_{C}$ . In the case of $N_{F}=N_{C}$, these
solutions are called local vortices and are expressed in terms of the moduli
corresponding to positions besides internal symmetries. On the other hand, the
solutions in case of $N_{F}>N_{C}$ are called semilocal vortices and require
the moduli corresponding not only to the position but also to the size and orientation.

Provided that $z\equiv x_{1}+ix_{2}$ is a complex coordinate for the $2$
dimensional space $R^{2}$, then the solutions of the BPS equations are
determined in general by the $N_{C}\times N_{F}$ matrix $H_{0}\left(
z\right)  $ which has elements consisting of holomorphic functions of $z$
\cite{Eto}. Here, $H_{0}$ is usually called a moduli matrix for the vortices
and we can represent any solution of BPS equations by means of $H_{0}$ as
follows. Let us introduce a $N_{C}\times N_{C}$ invertible matrix $S\left(
z,\bar{z}\right)  \in GL\left(  N_{C},\mathbb{C}\right)  $ and consider a
gauge invariant quantity defined by $\Omega\left(  z,\bar{z}\right)  \equiv
S\left(  z,\bar{z}\right)  S^{\dag}\left(  z,\bar{z}\right)  $. Then the Higgs
and gauge fields should be written as
\begin{align}
H  &  =S^{-1}H_{0}\ ,\nonumber\\
A_{1}+iA_{2}  &  =2S^{-1}\bar{\partial}_{z}S\ .
\end{align}
Actually, the first set of BPS equations could be solved for arbitrary $S$ on
account of these relations. And the second set of the BPS equations is written
in the form of%
\begin{equation}
\partial_{z}\left(  \Omega^{-1}\bar{\partial}_{z}\Omega\right)  =\dfrac{1}%
{2}\left(  \Omega^{-1}H_{0}H_{0}^{\dag}-c\boldsymbol{1}_{N_{C}}\right)  \ .
\end{equation}
This equation is called \emph{master equation} for the vortices and has a
unique solution $\Omega$ for any given $H_{0}$. Here, $S$ is determined except
for the gauge degrees of freedom and some ambiguities of decomposition. As a
result, we can find $H$ and $A_{i}$ which solve the BPS equation on account of
the relations given above.

Let us consider the case of local vortex with $N_{F}=N_{C}\equiv N$. The
moduli matrix $H_{0}\left(  z\right)  $ becomes a $N\times N$ matrix and it
can be shown that the vortex number is given by \cite{Eto}%
\begin{equation}
k=\dfrac{1}{2\pi}\operatorname{Im}\oint dz\partial_{z}\log\left(  \det
H_{0}\right)  \ .
\end{equation}
This representation for the topological charge makes it clear that $H_{0}$
behaves like $\det H_{0}\sim z^{k}$ \ at the spacial infinity $\left\vert
x\right\vert \rightarrow\infty$ . This agrees with the fact that the
dimensions of the vortex moduli space is equal to $2kN$ known from the index
theorem. It is known that the moduli space is constructed by the method which
is called K\"{a}hler quotient, and is represented as
\begin{equation}
\left\{  \boldsymbol{Z},\boldsymbol{\Psi}\right\}  //GL\left(  k,\mathbb{C}%
\right)  \simeq\left\{  \left(  Z,\psi\right)  |\left[  Z^{\dag},Z\right]
+\psi^{\dag}\psi\propto\boldsymbol{1}_{k}\right\}  /U\left(  k\right)  ,
\end{equation}
where $Z$ and $\psi$ are $k\times k$ and $N\times k$ matrices respectively
\cite{Tong,Eto}. This method to construct the vortex moduli extremely
resembles the ADHM method to construct the instanton moduli and it is called
half ADHM. It has not been clear why the moduli spaces of the vortex and the
instanton are constructed by such similar methods. In the following sections,
we will show that the non-Abelian vortex is equivalent to the instanton in
$R^{2}\times Z_{2}.$

\section{Gauge theory in noncommutative discrete space}

In order to make transparent the construction of the theory, we shall survey
the method of representing the gauge theory in the noncommutative space in
term of differential forms for the matrices. We propose that the gauge field
is an extended differential $(p+q)$-form on $M\times Z_{2}$ space consisting
of the $(p,q)$-forms on $M$ and $Z_{2}$ respectively.

Let us consider discrete two point space $Z_{2}$ which has noncommutative
nature in differential
calculus\cite{Connes,Coquereaux,Morita,Varilly,Teo:1997cn}. We employ matrices
as machineries representing such a structure and consider $2\times2\ $matrices
as$\ $differential forms in $Z_{2}$ space. Then we introduce $Z_{2}-$grading
corresponding to the parity with respect to the degrees of differential forms.
Where, the matrices with diagonal elements have even parity and the matrices
with anti-diagonal elements have odd parity.

In general, $2\times2\ $matrix
\begin{equation}
a=\left(
\begin{array}
[c]{cc}%
a_{11} & a_{12}\\
a_{21} & a_{22}%
\end{array}
\right)
\end{equation}
should be interpreted as a mixed differential form consisting of different
degrees of forms, which could be decomposed as $a=a_{e}+a_{o}$,
\begin{equation}
a_{e}=\left(
\begin{array}
[c]{cc}%
a_{11} & 0\\
0 & a_{22}%
\end{array}
\right)  ,\ a_{o}=\left(
\begin{array}
[c]{cc}%
0 & a_{12}\\
a_{21} & 0
\end{array}
\right)  \ .
\end{equation}
We represent the$\ Z_{2}-$parities of forms as $\left[  a_{e}\right]  =0\ $and
$\left[  a_{o}\right]  =1$ for even and odd matrices respectively. The wedge
product among the differential forms is to be assumed as that of the matrices.

$\ $The exterior derivative operator $d$ acting on the differential forms in
$Z_{2}$ space is defined as%
\begin{equation}
d=i\left[  \eta,\ \ \right\}  \ ,
\end{equation}
where $\left[  \alpha,\beta\right\}  $ is the graded commutator representing%
\begin{equation}
\left[  \alpha,\beta\right\}  =\alpha\beta-\left(  -1\right)  ^{\left[
\alpha\right]  \left[  \beta\right]  }\beta\alpha\
\end{equation}
and $\eta$ is an odd parity matrix with the property%
\begin{equation}
\eta^{2}=\boldsymbol{1\ .}%
\end{equation}
Then, the action of $d$ on arbitrary matrix differential form $\alpha$ is
\begin{equation}
d\alpha=i\left[  \eta,\alpha\right\}  =i\left(  \eta\alpha-\left(  -1\right)
^{\left[  \alpha\right]  }\alpha\eta\right)  \ .
\end{equation}
This leads to the result%
\begin{equation}
d^{2}\alpha=-\left[  \eta,\left[  \eta,\alpha\right\}  \right\}  =\dfrac{1}%
{2}\left[  \alpha,\left[  \eta,\eta\right\}  \right\}  =0\ ,
\end{equation}
where we use the relation
\begin{equation}
2\left[  \eta,\left[  \eta,\alpha\right\}  \right\}  +\left[  \alpha,\left[
\eta,\eta\right\}  \right\}  =0
\end{equation}
derived from the graded Jacobi identity%
\begin{equation}
\left(  -1\right)  ^{\left[  A\right]  \left[  C\right]  }\left[  A,\left[
B,C\right\}  \right\}  +\left(  -1\right)  ^{\left[  A\right]  \left[
B\right]  }\left[  B,\left[  C,A\right\}  \right\}  +\left(  -1\right)
^{\left[  C\right]  \left[  B\right]  }\left[  C,\left[  A,B\right\}
\right\}  =0
\end{equation}
and the fact that $\eta^{2}=\boldsymbol{1}$. Then we can consider $d$ itself
as nilpotent%
\begin{equation}
d^{2}=0\ ,
\end{equation}
which means that $d$ plays a role of an exterior derivative operator as it is
a linear operator with odd parity and has the nilpotency. The graded Leipniz's
rule
\begin{equation}
d\left(  \alpha\wedge\beta\right)  =d\alpha\wedge\beta+\left(  -1\right)
^{\left[  \alpha\right]  }\alpha\wedge d\beta
\end{equation}
is one of the desirable properties.

Although we may employ the form of
\begin{equation}
\eta=\eta_{\gamma}=\cos\gamma\tau_{1}+\sin\gamma\tau_{2}=\left(
\begin{array}
[c]{cc}%
0 & e^{-i\gamma}\\
e^{i\gamma} & 0
\end{array}
\right)  \ ,
\end{equation}
as $\eta$ in general, it is convenient to adopt
\begin{equation}
\eta=\eta_{0}=\left(
\begin{array}
[c]{cc}%
0 & 1\\
1 & 0
\end{array}
\right)  =\tau_{1}%
\end{equation}
to make calculations clear without loss of generality. We shall follow this
definition in this paper for simplicity, then we see that the action of $d$ on
the matrix $a$ results%
\begin{equation}
da=i\left[  \eta,a_{e}\right]  +i\left\{  \eta,a_{o}\right\}  =i\left(
\begin{array}
[c]{cc}%
a_{21}+a_{12} & a_{22}-a_{11}\\
a_{11}-a_{22} & a_{21}+a_{12}%
\end{array}
\right)  \ .
\end{equation}

Let us consider the differential forms in an extended space $M\times Z_{2}$
with $M$ an ordinary manifold. These are represented by $2\times2\ $matrices
whose elements are consisting of differential forms in $M$. For example, if we
consider the extended differential forms%
\begin{equation}
\mathcal{M}=\left(
\begin{array}
[c]{cc}%
A & C\\
D & B
\end{array}
\right)  ,\ \mathcal{M}^{\prime}=\left(
\begin{array}
[c]{cc}%
A^{\prime} & C^{\prime}\\
D^{\prime} & B^{\prime}%
\end{array}
\right)  ,
\end{equation}
then the wedge product of these is a product of two matrices accounting the
signature of grading as follows%
\begin{equation}
\mathcal{M\wedge M}^{\prime}=\left(
\begin{array}
[c]{cc}%
A\wedge A^{\prime}+\left(  -1\right)  ^{\left[  C\right]  }C\wedge D^{\prime}
& \left(  -1\right)  ^{\left[  A\right]  }A\wedge C^{\prime}+C\wedge
B^{\prime}\\
D\wedge A^{\prime}+\left(  -1\right)  ^{\left[  B\right]  }B\wedge D^{\prime}
& \left(  -1\right)  ^{\left[  D\right]  }D\wedge C^{\prime}+B\wedge
B^{\prime}%
\end{array}
\right)  \ .
\end{equation}
Here, $\left[  A\right]  $ stands for the $Z_{2}-$parity of degree of the
differential form $A$. The rule for the wedge product given above means that
$Z_{2}-$parity of the differential forms in $M$ should be identified with that
of the $2\times2\ $matrices in $Z_{2}$.

The arbitrary matrix $\mathcal{M}$ is represented as%
\begin{equation}
\mathcal{M}=e_{ij}\otimes\mathcal{M}_{ij}\ ,
\end{equation}
provided that the basis of $2\times2\ $matrices are assigned as%
\begin{equation}
e_{00}=\left(
\begin{array}
[c]{cc}%
1 & 0\\
0 & 0
\end{array}
\right)  ,\ e_{01}=\left(
\begin{array}
[c]{cc}%
0 & 1\\
0 & 0
\end{array}
\right)  ,\ e_{10}=\left(
\begin{array}
[c]{cc}%
0 & 0\\
1 & 0
\end{array}
\right)  ,\ e_{11}=\left(
\begin{array}
[c]{cc}%
0 & 0\\
0 & 1
\end{array}
\right)  \ ,
\end{equation}
with the $Z_{2}-$parity $\left[  e_{00}\right]  =\left[  e_{11}\right]
=0\ \left(  \text{even}\right)  ,\ \left[  e_{01}\right]  =\left[
e_{10}\right]  =1\ \left(  \text{odd}\right)  .$ Here we employ a
representation in terms of the direct product. And keep a convention that the
basis of $2\times2$ matrix forms should be located on the left of the ordinary
differential forms. Then the wedge product of the two matrix differential
forms $\mathcal{M}$ and $\mathcal{M}^{\prime}$ is calculated as
\begin{align}
\mathcal{M\wedge M}^{\prime}  &  =\left(  e_{ij}\otimes\mathcal{M}%
_{ij}\right)  \wedge\left(  e_{kl}\otimes\mathcal{M}_{kl}^{\prime}\right)
\nonumber\\
&  =\left(  e_{ij}\ e_{kl}\right)  \otimes\left(  \left(  -1\right)  ^{\left[
\mathcal{M}_{ij}\right]  \left[  e_{kl}\right]  }\mathcal{M}_{ij}%
\wedge\mathcal{M}_{kl}^{\prime}\right) \nonumber\\
&  =\left(  \delta_{jk}~e_{il}\right)  \otimes\left(  \left(  -1\right)
^{\left[  \mathcal{M}_{ij}\right]  \left[  e_{kl}\right]  }\mathcal{M}%
_{ij}\wedge\mathcal{M}_{kl}^{\prime}\right) \nonumber\\
&  =e_{il}\otimes\left(  \left(  -1\right)  ^{\left[  \mathcal{M}_{ij}\right]
\left[  e_{jl}\right]  }\mathcal{M}_{ij}\wedge\mathcal{M}_{jl}^{\prime
}\right)  \ .
\end{align}
It is worth to be remarked that the same sign rule
\begin{equation}
\alpha\mathcal{\wedge\beta=}\left(  -1\right)  ^{\left[  \alpha\right]
\left[  \mathcal{\beta}\right]  }\mathcal{\beta\wedge\alpha}%
\end{equation}
should be applied when we exchange the order between the basis of matrix
differential forms and the ordinary differential forms. There exists similar
idea to define the hermitian conjugate of the graded matrix differential form.
It is defined by%
\begin{equation}
\ \mathcal{M}^{\dag}=\left(
\begin{array}
[c]{cc}%
A^{\dag} & \left(  -1\right)  ^{\left[  D\right]  }D^{\dag}\\
\left(  -1\right)  ^{\left[  C\right]  }C^{\dag} & B^{\dag}%
\end{array}
\right)
\end{equation}
for $\mathcal{M}$ given above.

To make the difference clear, let us use the symbols $d_{H}$ and $d_{V}$ to
represent the exterior derivative operators on $M\left(  \text{horizontal}%
\right)  $ and $Z_{2}\left(  \text{vertical}\right)  $ spaces respectively.
That is to say, $d_{H}$ represents the exterior derivative operator acting on
the ordinary differential forms and $d_{V}$ represents the exterior derivative
on the matrix differential forms. If we identify the $Z_{2}-$parity of the
matrix differential forms with that of the ordinary differential forms,\ then
we can agree that the operators $d_{H}$ and $d_{V}$ anti-commute with each
other
\begin{equation}
d_{H}d_{V}=-d_{V}d_{H}\ .
\end{equation}
Thus, we consider the exterior derivative operator acting on the generalized
differential form in $M\times Z_{2}$
\begin{equation}
\boldsymbol{d}=d_{H}+d_{V}%
\end{equation}
which satisfies the nilpotency as%
\begin{equation}
\boldsymbol{d}^{2}=d_{H}^{2}+d_{H}d_{V}+d_{V}d_{H}+d_{V}^{2}=0\ .
\end{equation}
The explicit action of $\boldsymbol{d}$ on the matrix differential form
$\mathcal{M}$ is
\begin{align}
\boldsymbol{d}\mathcal{M}  &  =d_{H}\mathcal{M}+d_{V}\mathcal{M}\nonumber\\
&  =\left(
\begin{array}
[c]{cc}%
dA & -dC\\
-dD & dB
\end{array}
\right)  +i\left[  \eta,\left(
\begin{array}
[c]{cc}%
A & 0\\
0 & B
\end{array}
\right)  \right]  +i\left\{  \eta,\left(
\begin{array}
[c]{cc}%
0 & C\\
D & 0
\end{array}
\right)  \right\} \nonumber\\
&  =\left(
\begin{array}
[c]{cc}%
dA+i\left(  C+D\right)  & -dC-i\left(  A-B\right) \\
-dD+i\left(  A-B\right)  & dB+i\left(  C+D\right)
\end{array}
\right)  ,
\end{align}
where we simply write $d$ in place of $d_{H}$ when its role is manifest.

Actually, we can assign not only the $Z_{2}-$parity but also the degree to
matrix differential forms in $Z_{2}$ space. In general, $2\times2$ matrix can
be expanded in terms of the basis $\left(  \boldsymbol{1},\tau_{1},\tau
_{2},\tau_{3}\right)  $ with Pauli matrices $\tau_{i}$ 's. These bases are
classified into the even basis $\left(  1,\tau_{3}\right)  $ and odd one
$\left(  \tau_{1},\tau_{2}\right)  .$ It is natural to consider $\mathbf{1}$
as a basis of 0-form and $\left(  \tau_{1},\tau_{2}\right)  $ as basis of
1-form according to their $Z_{2}-$parity. The basis of 1-form is often written
as $\left(  \theta_{1},\theta_{2}\right)  $. Then the basis of 2-form can be
obtained by
\begin{equation}
\theta_{1}\wedge\theta_{2}=\tau_{1}\tau_{2}=i\tau_{3}\ .
\end{equation}
The $2\times2$ matrix $\mathcal{M}$ is expanded as
\begin{equation}
\mathcal{M}=\left(
\begin{array}
[c]{cc}%
A & C\\
D & B
\end{array}
\right)  =\boldsymbol{1}\otimes\frac{A+B}{2}+\tau_{1}\otimes\frac{C+D}{2}%
+\tau_{2}\otimes\frac{i\left(  C-D\right)  }{2}+\tau_{3}\otimes\frac{A-B}%
{2}\ ,
\end{equation}
then we can interpret these terms as different degrees of forms in $Z_{2}$ space.

We can define a gauge theory in $M\times Z_{2}$ in terms of the matrix
differential forms. Let us consider a gauge field as a connection 1-form in
$M\times Z_{2}$ space in the form of%
\begin{equation}
\mathcal{A}=\left(
\begin{array}
[c]{cc}%
L & i\varphi\\
i\varphi^{\dag} & R
\end{array}
\right)  ,
\end{equation}
where $L,R$ and $\varphi$ are Lie-algebra valued $1$-forms and $0$-form on $M$
respectively. It should be remarked that $L$ and $R$ are anti-hermitian and
$\varphi$ is complex. Let our model be that consisting of $L,R,\varphi$ with
value on $N\times N$ matrices. It means that the model has the $U\left(
N\right)  _{L}\times U\left(  N\right)  _{R}$ gauge symmetry. The connection
form is expanded as
\begin{equation}
\mathcal{A}=\mathcal{A}^{\left(  1,0\right)  }\mathcal{+A}^{\left(
1,2\right)  }\mathcal{+A}^{\left(  0,1\right)  },
\end{equation}
where%
\begin{align}
\mathcal{A}^{\left(  1,0\right)  }  &  =\boldsymbol{1}\otimes\frac{L+R}%
{2},\nonumber\\
\mathcal{A}^{\left(  1,2\right)  }  &  =\tau_{3}\otimes\frac{L-R}%
{2},\nonumber\\
\mathcal{A}^{\left(  0,1\right)  }  &  =\tau_{1}\otimes\frac{\varphi
+\varphi^{\dag}}{2}+\tau_{2}\otimes\frac{i\left(  \varphi-\varphi^{\dag
}\right)  }{2}.
\end{align}
We use the notation $\mathcal{A}^{\left(  p,q\right)  }$ to represent a form
that behaves itself as a $p$-form in $M$ and as a matrix $q$-form in $Z_{2}%
\ $, that is, a $\left(  p+q\right)  $-form in $M\times Z_{2}$ space as a whole.

The field strength $\mathcal{F}$ derived from the gauge field $\mathcal{A}$ is
defined by
\begin{equation}
\mathcal{F}=\boldsymbol{d}\mathcal{A}+\mathcal{A}\wedge\mathcal{A}%
\end{equation}
or
\begin{align}
\mathcal{F}  &  =\left(
\begin{array}
[c]{cc}%
dL+L\wedge L-\varphi\varphi^{\dag}-\left(  \varphi+\varphi^{\dag}\right)  &
-i\left(  d\varphi+\left(  L\varphi-\varphi R\right)  +\left(  L-R\right)
\right) \\
-id\left(  d\varphi^{\dag}-\left(  \varphi^{\dag}L-R\varphi^{\dag}\right)
-\left(  L-R\right)  \right)  & dR+R\wedge R-\varphi^{\dag}\varphi-\left(
\varphi+\varphi^{\dag}\right)
\end{array}
\right) \nonumber\\
&  =\left(
\begin{array}
[c]{cc}%
F^{L}-W^{L} & -iD\phi\\
-i\left(  D\phi\right)  ^{\dag} & F^{R}-W^{R}%
\end{array}
\right)
\end{align}
in components. Where, $F^{L}\ $and $F^{R}$ are field strengths of the gauge
fields $L$ and $R$ on the manifold $M$,%
\begin{align}
F^{L}  &  =dL+L\wedge L\ ,\nonumber\\
F^{R}  &  =dR+R\wedge R\ .
\end{align}
$D$ is a covariant derivative with respect to both $L\ $and $R\ $,
\begin{equation}
D\phi=d\phi+L\phi-\phi R\ ,
\end{equation}
provided that $L$ and $R$ act from the left and right respectively. $\phi$ is
defined by%
\begin{equation}
\phi=\varphi+\boldsymbol{1}_{N}\ ,
\end{equation}
then $W^{L}$ and $W^{R}$ are defined by%
\begin{align}
W^{L}\left(  \phi\right)   &  =\left(  \varphi+\boldsymbol{1}_{N}\right)
\left(  \varphi^{\dag}+\boldsymbol{1}_{N}\right)  -\boldsymbol{1}_{N}=\phi
\phi^{\dag}-\boldsymbol{1}_{N}\ ,\nonumber\\
W^{R}\left(  \phi\right)   &  =\left(  \varphi^{\dag}+\boldsymbol{1}%
_{N}\right)  \left(  \varphi+\boldsymbol{1}_{N}\right)  -\boldsymbol{1}%
_{N}=\phi^{\dag}\phi-\boldsymbol{1}_{N}\ .
\end{align}
As we shall see in the following section,\ Higgs potential $V\left(
\phi\right)  $ can be given by $W^{L}$ and $W^{R}$ as
\begin{equation}
V\left(  \phi\right)  =\mathrm{Tr}\left(  W^{L}\right)  ^{2}=\mathrm{Tr}%
\left(  W^{R}\right)  ^{2}=\mathrm{Tr}\left(  \phi\phi^{\dag}-\boldsymbol{1}%
_{N}\right)  ^{2}=\mathrm{Tr}\left(  \phi^{\dag}\phi-\boldsymbol{1}%
_{N}\right)  ^{2}\ ,
\end{equation}
where $\mathrm{Tr}$ is a trace over the representation matrix of Lie-algebra.
It means that $\phi$ is the Higgs filed and that $\varphi$ is its fluctuation
around the vacuum expectation value $\boldsymbol{1}_{N}$.

There exists an appropriate definition of the Hodge dual $^{\ast}\mathcal{F}$
of $\mathcal{F}$ and a definition of volume integral of norm square of
$\mathcal{F}$ on $M\times Z_{2}$ space, that is%
\begin{equation}
\mathrm{Tr}\int_{M\times Z_{2}}\left\langle \mathcal{F},\mathcal{F}%
\right\rangle =\mathrm{Tr}\int_{M\times Z_{2}}\mathcal{F}\wedge^{\ast
}\mathcal{F\ }.
\end{equation}
The action of the gauge theory is nothing but this integral and we have
\begin{align}
S &  =\frac{1}{2}\mathrm{Tr}\int_{M\times Z_{2}}\mathcal{F}\wedge^{\ast
}\mathcal{F}\nonumber\\
&  \mathcal{=}\mathrm{Tr}\int_{M}\left(  \frac{1}{2}\left\vert F_{12}%
^{L}\right\vert ^{2}+\frac{1}{2}\left\vert F_{12}^{R}\right\vert ^{2}+\left(
\phi\phi^{\dag}-\boldsymbol{1}_{N}\right)  ^{2}+\left\vert D_{1}%
\phi\right\vert ^{2}+\left\vert D_{2}\phi\right\vert ^{2}\right)  dx^{1}\wedge
dx^{2}\wedge\cdots\wedge dx^{n}%
\end{align}
as a result. The concrete definition of Hodge duality is necessary to express
the Yang-Mills action by means of the inner product among the differential
forms on the noncommutative space. The derivation of the above action with the
definite form of Hodge duality will be explained in some detail in the next section.

This action is that of the Yang-Mills-Higgs model, which consists of the
kinetic terms of the Yang-Mills gauge field and the Higgs field. This means
that pure Yang-Mills gauge theory in $M\times Z_{2}$ is equivalent to the
Yang-Mills-Higgs theory in $M$ which is automatically incorporated in the
mechanism of spontaneous symmetry breaking in natural way.

Let us think of the case of $N=1$ that is abelian gauge theory for simplicity,
we see that the combination of $L+R$ becomes massive and $L-R$ remains
massless. That is to say, this is a model of the Higgs mechanism which breaks
a gauge symmetry $U\left(  1\right)  _{L}\times U\left(  1\right)  _{R}$ to
$U\left(  1\right)  _{L-R}$ . If we adapt this machinery to the standard
model, it would be suitable to assign $U\left(  2\right)  _{L}\times U\left(
1\right)  _{R}$ as a gauge group, with total trace free condition.

In this construction, we can understand that the Higgs field is included as a
kind of gauge field by generalizing the gauge theory to the space with
discrete and noncommutative geometry. Then this method leads to the Higgs
mechanism and spontaneous symmetry breaking naturally, which is nothing but
the gauge theory itself. There has been many explicit applications of this
idea to reconstruct standard model\cite{Connes,Coquereaux,Morita}.

\section{Non-Abelian vortex on $R^{2}$ as instanton on $R^{2}\times Z_{2}$}

The idea of Hodge duality plays a crucial role in understanding the relation
of instantons and vortices. Actually, to interpret the non-Abelian vortex on
$R^{2}$ as an instanton on $R^{2}\times Z_{2}$, the concept of Hodge duality
for the matrix differential forms on the noncommutative space has to be
defined. This concept, which seems to be not necessarily well defined in the
literature, is indispensable.

In this section, we have worked out the concrete definition of Hodge duality.
Based on this, we have expressed the Yang-Mills action, which we have
explained in the previous section, on the noncommutative space in terms of the
Hodge dual operation. Furthermore under the operation of the Hodge dual, we
describe the instanton equation on $R^{2}\times Z_{2}$, and reveal the fact
that it is equivalent to the vortex equation on $R^{2}$.

The general $p$-form in the ordinary manifold $M$ can be written as%
\begin{equation}
\alpha=\frac{1}{p!}\alpha_{\mu_{1}\mu_{2}\cdots\mu_{p}}dx^{\mu_{1}}\wedge
dx^{\mu_{2}}\wedge\cdots\wedge dx^{\mu_{p}}%
\end{equation}
in terms of the basis $dx^{\mu}$ of the covariant vectors which span the
cotangent vector bundle $T^{\ast}\left(  M\right)  $. One can also introduce
the dual basis $\dfrac{\partial}{\partial x^{\mu}}\ $to $dx^{\mu}$ , that is,
the basis of the contravariant vectors which span the tangent vector bundle
$T\left(  M\right)  $ of the manifold $M$. The inner product among the basis
and dual one is defined to satisfy the relation%
\begin{equation}
\left\langle dx^{\mu},\dfrac{\partial}{\partial x^{\nu}}\right\rangle
=\delta_{\nu}^{\mu}\ .
\end{equation}

The Hodge dual operation $\ast$ is defined so as to transfer the inner product
of $\alpha$ and $\beta$ to the wedge product of $\alpha$ and $^{\ast}\!\beta$
\begin{equation}
\left\langle \alpha,\beta\right\rangle dx^{1}\wedge dx^{2}\wedge\cdots\wedge
dx^{n}=\alpha\wedge\!\left.  {}\right.  ^{\ast}\!\beta\ .
\end{equation}
Let us define the explicit correspondence in terms of the Hodge dual of the
forms in $R^{2}$ as follows. The Hodge dual of the $2$-form $F$, $1$-form $V$
and $0$-form $W$,
\begin{align}
F  &  =\frac{1}{2}F_{ij}dx^{i}\wedge dx^{j}\ ,\nonumber\\
V  &  =V_{i}dx^{i}\ ,
\end{align}
are given by the $0$-form $^{\ast}F$, $1$-form $^{\ast}V$ and $2$-form
$^{\ast}W$,%
\begin{align}
^{\ast}F  &  =\frac{1}{2}\varepsilon_{ji}F_{ij}=-F_{12}\ ,\nonumber\\
^{\ast}V  &  =\varepsilon_{ij}V_{j}dx^{i}\ ,\nonumber\\
^{\ast}W  &  =\frac{1}{2}\varepsilon_{ij}Wdx^{i}\wedge dx^{j}\ ,
\end{align}
respectively. As a result of the definition given above, we see that $\left.
{}\right.  ^{\ast\ast}=-1$ for the forms of arbitrary degrees. This means that
our definition results in $\left.  {}\right.  ^{\ast\ast}=1$ when it is
extended to the case of 4 dimensional space $R^{4}$ with Euclidean signature.

As the eigen values of the Hodge duality operator $\ast$ in $R^{4}$ are $\pm
1$, we can define the (anti-)selfdual $2$-form $F_{+}$ $\left(  F_{-}\right)
$ by\
\begin{equation}
F_{\pm}=F\pm\left.  {}\right.  ^{\ast}F
\end{equation}
which are the eigen states of the operator $\ast$
\begin{equation}
^{\ast}F_{\pm}=\pm F_{\pm}%
\end{equation}
with respect to an arbitrary $2$-form $F$ in $R^{4}$ .

On the other hand, the eigen values of the Hodge duality operator $\ast$ in
$R^{2}$ are $\pm i$ . Then we can define the (anti-)selfdual 1-form $V_{+}$
$\left(  V_{-}\right)  $ by%
\begin{equation}
V_{\pm}=V\mp i^{\ast}V
\end{equation}
which are the eigen states of the operator $\ast$%
\begin{equation}
^{\ast}V_{\pm}=\pm iV_{\pm}%
\end{equation}
with respect to an arbitrary $1$-form $V$ in $R^{2}$ . $V_{\pm}$ has the
components $V_{\pm}=\left(  \left(  V_{\pm}\right)  _{1},\left(  V_{\pm
}\right)  _{2}\right)  $
\begin{align}
\left(  V_{\pm}\right)  _{1}  &  =V_{1}\mp iV_{2}\ ,\nonumber\\
\left(  V_{\pm}\right)  _{2}  &  =V_{2}\pm iV_{1}=\pm i\left(  V_{1}\mp
iV_{2}\right)  =\pm i\left(  V_{\pm}\right)  _{1}\ ,
\end{align}
with $V=\left(  V_{1},V_{2}\right)  $ .

Let us consider the case of the matrix differential forms. As the basis
$\theta^{a}$ of the covariant vectors are represented by the matrices, the
dual basis or the contravariant vectors $e_{a}$ which satisfy the relation
\begin{equation}
\left\langle \theta^{a},e_{b}\right\rangle =\delta_{b}^{a}\ ,
\end{equation}
are also represented by matrices. We can see that this inner product is the
normalized trace of such matrices. Actually in case of matrix differential
forms in $Z_{2}$, as we employ the convention that the basis are represented
by $\theta^{1}=\tau_{1},\ \theta^{2}=\tau_{2}$, the dual basis have the same
forms as themselves,
\begin{equation}
e_{1}=\tau_{1},\ e_{2}=\tau_{2}.
\end{equation}
In this representation, the Hodge dual operation is equal to the
multiplication by $i\tau_{3}$ . This operation maps the basis of the matrix
$0,1,2$-forms, $\left\{  1,\left(  \theta^{1},\theta^{2}\right)  ,\theta
^{1}\wedge\theta^{2}\right\}  $ or $\left\{  1,\left(  \tau_{1},\tau
_{2}\right)  ,i\tau_{3}\right\}  $ to the $2,1,0$-forms, $\left\{  \theta
^{1}\wedge\theta^{2},\left(  \theta^{2},-\theta^{1}\right)  ,-1\right\}  \ $or
$\left\{  i\tau_{3},\left(  \tau_{2},-\tau_{1}\right)  ,-1\right\}  .$ These
results coincide with that of the operation $\ast$ in $R^{2}$ described above.
Thus we can see that the Hodge dual in $R^{2}\times Z_{2}$ as a
four-dimensional space is performed by the duality operation in $R^{2}$ and
$Z_{2}$ at the same time. As a result, we obtain real values $\pm1$ as the
eigen values of the Hodge duality operator in $R^{2}\times Z_{2},\ $although
those values are imaginary $\pm i$ in $R^{2}$ and $Z_{2}.$

Let us consider the pure Yang-Mills action in $R^{2}\times Z_{2}$. The Hodge
dual of field strength $2$-form for the gauge field in $R^{2}\times Z_{2}$ is
given by%
\begin{equation}
^{\ast}\mathcal{F}\equiv i\tau_{3}\mathcal{F}\left(  ^{\ast}\right)  \ .
\end{equation}
Here, we mean $\mathcal{F}\left(  ^{\ast}\right)  $ the Hodge dual with
respect to the forms in $R^{2}$ as the components of $2\times2$ matrix
$\mathcal{F}$ , whereas the Hodge dual with respect to the matrix differential
forms in $Z_{2}$ is represented by the multiplication with $i\tau_{3}$ .

We could obtain the action for pure Yang-Mills theory on $R^{2}\times Z_{2}$
by integration of the Lagrangian%
\begin{equation}
\mathcal{L}=\dfrac{1}{2}\mathrm{Tr}\left\langle \mathcal{F},\mathcal{F}%
\right\rangle =\dfrac{1}{2}\mathrm{Tr}\mathcal{F}\wedge^{\ast}\mathcal{F}%
\end{equation}
over the volume of $R^{2}\times Z_{2}$. Although different degrees of forms
coexist in $\mathcal{F}\wedge^{\ast}\mathcal{F}$, we would pick up the volume
form $dx^{1}\wedge dx^{2}\wedge\theta^{1}\wedge\theta^{2}$ out of it.
Accounting the fact that the volume form $\theta^{1}\wedge\theta^{2}$ of the
$Z_{2}$ space is equal to $i\tau_{3}$ in our convention, we have to take a
trace of $\mathcal{L}$ after multiplying by $-i\tau_{3}$ as a $2\times2$
matrix, in order to pick up the volume form. Thus the volume integral
$\int_{Z_{2}}\left(  \mathcal{\ }\right)  $ over the $Z_{2}$ space is
equivalent to $-\frac{i}{2}$ $\mathrm{Tr}_{Z_{2}}\tau_{3}\left(  \ \right)  $
, where $\mathrm{Tr}_{Z_{2}}$ represents a trace with respect to $2\times2$
matrix as a differential form in $Z_{2}$ space. It would be clear that the
volume integral $\int_{R^{2}}\left(  \mathcal{\ }\right)  $ over the $R^{2}$
leaves $2$-forms. As a results, we have the action in the form of%
\begin{align}
S  &  =\mathrm{Tr}\int_{R^{2}\times Z_{2}}\mathcal{F}\wedge^{\ast}%
\mathcal{F}\nonumber\\
&  =\dfrac{1}{2}\mathrm{Tr}\int_{R^{2}}\mathrm{Tr}_{Z_{2}}\tau_{3}%
\mathcal{F}\wedge\tau_{3}\mathcal{F}\left(  ^{\ast}\right) \nonumber\\
&  =\dfrac{1}{2}\mathrm{Tr}\int_{R^{2}}\mathrm{Tr}_{Z_{2}}\tau_{3}\left(
\begin{array}
[c]{cc}%
F^{L}-W^{L} & -iD\phi\\
-i\left(  D\phi\right)  ^{\dag} & F^{R}-W^{R}%
\end{array}
\right)  \wedge\left(
\begin{array}
[c]{cc}%
^{\ast}\left(  F^{L}-W^{L}\right)  & -i^{\ast}\left(  D\phi\right) \\
i^{\ast}\left(  D\phi\right)  ^{\dag} & -^{\ast}\left(  F^{R}-W^{R}\right)
\end{array}
\right) \nonumber\\
&  \mathcal{=}\mathrm{Tr}\int_{R^{2}}\left(  \frac{1}{2}\left\vert F_{12}%
^{L}\right\vert ^{2}+\frac{1}{2}\left\vert F_{12}^{R}\right\vert ^{2}+\left(
\phi\phi^{\dag}-\boldsymbol{1}_{N}\right)  ^{2}+\left\vert D_{1}%
\phi\right\vert ^{2}+\left\vert D_{2}\phi\right\vert ^{2}\right)  dx^{1}\wedge
dx^{2}.
\end{align}
Thus we can confirm the fact that the action for pure Yang-Mills theory in
$R^{2}\times Z_{2}$ is equivalent to the action for Yang-Mills-Higgs theory in
$R^{2}.$

We consider a model with $U\left(  N\right)  _{L}\times U\left(  N\right)
_{R}$ gauge symmetry, where the fields $L,R$ and $\varphi$ are $N\times N$
matrices in general. In order to obtain a model for the non-Abelian vortex
considered in ref's \cite{Tong,Eto}, it would be required to make an
appropriate reduction, that is to restrict the gauge group to $U\left(
N\right)  _{L}\times U\left(  1\right)  _{R}$ . Then we combine $U\left(
1\right)  _{R}$ with $U\left(  1\right)  _{L}$, that is a subgroup of
$U\left(  N\right)  _{L}$, to obtain $U\left(  1\right)  _{L-R}$ and $U\left(
1\right)  _{L+R}$, the latter of which is decoupled from the other fields. If
we discard the decoupled $U\left(  1\right)  $, we have a model with $U\left(
N\right)  $ gauge symmetry, which describes local vortex. Our general model is
considered as that with the extension to have a local flavor symmetry, which
should be frozen to become a global one.

Now, we shall show that the instanton on $R^{2}\times Z_{2}$ is equivalent to
the vortex on $R^{2}.$ The field strength \textquotedblleft$2$-form"
$\mathcal{F}$ of the gauge field on $\mathcal{M}_{4}=R^{2}\times Z_{2}$ can be
decomposed as
\begin{equation}
\mathcal{F=F}^{\left(  0,0\right)  }+\mathcal{F}^{\left(  2,0\right)
}+\mathcal{F}^{\left(  1,1\right)  }+\mathcal{F}^{\left(  0,2\right)
}+\mathcal{F}^{\left(  2,2\right)  },
\end{equation}
provided that the basis of $0,1,2$-forms on $Z_{2}$ are $\boldsymbol{1}%
,\left(  \tau_{1},\tau_{2}\right)  ,i\tau_{3}$ respectively, \ where%
\begin{align}
\mathcal{F}^{\left(  0,0\right)  }  &  =\mathbf{1}\otimes\left(  -\frac
{W^{L}+W^{R}}{2}\right)  ,\nonumber\\
\mathcal{F}^{\left(  2,0\right)  }  &  =\mathbf{1}\otimes\left(  \frac
{F^{L}+F^{R}}{2}\right)  ,\nonumber\\
\mathcal{F}^{\left(  1,1\right)  }  &  =-i\left(  \tau_{1}\otimes\left(
\frac{D\phi+\left(  D\phi\right)  ^{\dag}}{2}\right)  +\tau_{2}\otimes
i\left(  \frac{D\phi-\left(  D\phi\right)  ^{\dag}}{2}\right)  \right)
,\nonumber\\
\mathcal{F}^{\left(  0,2\right)  }  &  =i\tau_{3}\otimes\left(  -i\right)
\left(  -\frac{W^{L}-W^{R}}{2}\right)  ,\nonumber\\
\mathcal{F}^{\left(  2,2\right)  }  &  =i\tau_{3}\otimes\left(  -i\right)
\left(  \frac{F^{L}-F^{R}}{2}\right)  .
\end{align}
With respect to the total degrees, $\mathcal{F}$ should be understood as an
mixed form consisting of not only total $2$-form $\mathcal{F}^{\left(
2,0\right)  }+\mathcal{F}^{\left(  1,1\right)  }+\mathcal{F}^{\left(
0,2\right)  }$ but also $0$-form $\mathcal{F}^{\left(  0,0\right)  }$ and
$4$-form $\mathcal{F}^{\left(  2,2\right)  }.$

The Hodge operator $\ast$ in $R^{2}\times Z_{2}$ transfers the field strength
$\mathcal{\mathcal{F}}$ into its dual $^{\ast}\mathcal{F}\equiv i\tau
_{3}\mathcal{F}\left(  ^{\ast}\right)  $ according to the definition in the
previous section. As a result, the components of $\mathcal{\mathcal{F}}$ in
the above decomposition are transferred as%
\begin{equation}
\mathcal{\mathcal{F}}^{\left(  p,q\right)  }\rightarrow\mathcal{\ ^{\ast
}\mathcal{F}}^{\left(  2-p,2-q\right)  }.
\end{equation}
Then we can see that there is a correspondence not only between $2$-form and
dual $2$-form but also between $0$-form and $4$-form. The instanton equation
for Yang-Mills gauge field in $4$ dimensions is nothing but a requirement of
(anti-)selfduality for the field strength $2$-form,
\begin{equation}
^{\ast}\mathcal{F=\pm F}.
\end{equation}
For the case of the gauge field in $R^{2}\times Z_{2}$, the (anti-)selfdual
Yang-Mills equation means the correspondence of the form
\begin{equation}
^{\ast}\mathcal{F}^{\left(  p,q\right)  }=\mathcal{\pm F}^{\left(
2-p,2-q\right)  },
\end{equation}
that is,%
\begin{equation}
i\tau_{3}\mathcal{F}^{\left(  p,q\right)  }\left(  ^{\ast}\right)
=\mathcal{\pm F}^{\left(  2-p,2-q\right)  },
\end{equation}
on account of nature of the Hodge operator for the matrix differential forms.
This equation is decomposed as follows in terms of the differential forms in
$R^{2}$%
\begin{equation}
\ ^{\ast}\left(  \frac{F^{L}+F^{R}}{2}\right)  =\pm\left(  -i\right)  \left(
-\frac{W^{L}-W^{R}}{2}\right)
\end{equation}
for $\left(  p,q\right)  =\left(  2,0\right)  $ or $\left(  0,2\right)  ,$%
\begin{equation}
\ ^{\ast}\left(  -\frac{W^{L}+W^{R}}{2}\right)  =\pm\left(  -i\right)  \left(
\frac{F^{L}-F^{R}}{2}\right)
\end{equation}
for $\left(  p,q\right)  =\left(  0,0\right)  $ or $\left(  2,2\right)  ,$ and%
\begin{align}
&  \left(  i\tau_{3}\right)  \left(  \tau_{1}\otimes\ ^{\ast}\left(
\frac{D\phi+\left(  D\phi\right)  ^{\dag}}{2}\right)  +\tau_{2}\otimes
i\ ^{\ast}\left(  \frac{D\phi-\left(  D\phi\right)  ^{\dag}}{2}\right)
\right) \nonumber\\
&  =\pm\left(  \tau_{1}\otimes\left(  \frac{D\phi+\left(  D\phi\right)
^{\dag}}{2}\right)  +\tau_{2}\otimes i\left(  \frac{D\phi-\left(
D\phi\right)  ^{\dag}}{2}\right)  \right)
\end{align}
for $\left(  p,q\right)  =\left(  1,1\right)  $. As a result, we have the
(anti-)selfdual equations written as
\begin{align}
iF_{12}^{L}\pm W^{L}  &  =0,\nonumber\\
iF_{12}^{R}\mp W^{R}  &  =0,\nonumber\\
\left(  D_{1}\pm iD_{2}\right)  \phi &  =0,
\end{align}
in components.

These equations can also be obtained by considering \ on the equal footing the
differential forms of different nature. Suppose the basis of $1$-form in
\textquotedblleft4 dimensional" space $\mathcal{M}_{4}=R^{2}\times Z_{2}$ to
be $dx^{\mu}$ $\left(  \mu=1,2,3,4\right)  $. Let us consider indices 1,2 to
show ingredients of basis of the 1-form in 2 dimensional continuous space
$R^{2}$ and index 3,4 to show those in \textquotedblleft2 dimensional"
discrete space $Z_{2}.$ That means assigning%
\begin{equation}
\left(  dx^{3},dx^{4}\right)  =\left(  \tau^{1},\tau^{2}\right)
\end{equation}
besides the ordinary basis $\left(  dx^{1},dx^{2}\right)  $. On account of
these assignments, we can also obtain the (anti-)selfdual Yang-Mills equations
in the same way as the usual expression in terms of the components.

This is a BPS equation expressing the non-Abelian vortex, so we have shown
that an instanton equation in $R^{2}\times Z_{2}$ was none other than the
non-Abelian vortex equation in $R^{2}$. It can also be verified that the
instanton number or a 2nd Chern character in $R^{2}\times Z_{2}$ is just the
vortex number in $R^{2}$ as follows. Remember the volume integral over the
$Z_{2}$ is equal to the trace after multiplication with $-\frac{i}{2}\tau
_{3},$ then we have the relation%

\begin{equation}
-\mathrm{Tr}\int_{R^{2}\times Z_{2}}\mathcal{F}\wedge\mathcal{F=}i\int_{R^{2}%
}\frac{1}{2}\mathrm{Tr}_{Z_{2}}\tau_{3}\mathcal{F}\wedge\mathcal{F}%
=i\int_{R^{2}}\left(  \mathrm{Tr}F_{12}^{L}-\mathrm{Tr}F_{12}^{R}\right)
dx^{1}\wedge dx^{2},
\end{equation}
which means that instanton number on $R^{2}\times Z_{2}$ is given by the
difference between the vortex numbers of two gauge fields on $R^{2}.$ Actually
$\mathrm{Tr}F_{12}^{L}$ and $\mathrm{Tr}F_{12}^{R}$ have opposite sign, when
we consider non trivial vortex solutions. As a result, we have vortex number
as an instanton number in noncommutative discrete space.

\section{Discussion}

In this article, we have employed a matrix differential form to express
differential geometry of noncommutative discrete space. As has been described
in this work, the non-Abelian vortex in $R^{2}$ is equivalent to the instanton
on the $R^{2}\times Z_{2}.$ This suggests the possibility to constitute
non-Abelian vortex solution by the ADHM method. Actually, the moduli for the
non-Abelian vortices are described by the method that resembles ADHM which is
named \textquotedblleft half ADHM", though the explicit form of the solutions
is not decided yet. The relation between the instanton and the vortex that we
reported in this article shows possibility to clarify the reason why the
half-ADHM method works.

In the usual ADHM method for instanton\cite{Corrigan:1977ma}, we employ a
quaternionic variables $x$ as a coordinate of $4$ dimensional space $R^{4}$.
The ADHM data, that is, moduli parameters to describe instantons, are included
into the\ \textquotedblleft$0$ dimensional Dirac operator" $\nabla=Cx-D$ as
its coefficient matrices, $C$ and $D$. We should solve the \textquotedblleft
Dirac equation" $\nabla^{\dag}V=0,$ in order to determine the gauge connection
in the form of $A=V^{\dag}dV$ with $V$ which satisfies $V^{\dag}V=1$. The
condition for the field strength to be anti-selfdual is that $\nabla^{\dag
}\nabla$ should be an invertible matrix which consists of the real numbers
although $\nabla$ itself has quaternionic entities.

Although differential forms and the calculation rule among them are given in
the case of $R^{2}\times Z_{2}$ space, we do not know an exact expression for
the coordinates in this space. The operator $\nabla$ therefore is not yet
given which is a key issue to clarify how to construct the ADHM in this case.
It is suggested that there exists a mechanism analogous to ADHM even if the
exact expression of the coordinate is unidentified. For example, the gauge
field is given in the form of a kind of non-linear sigma models. If we
tentatively assume that the extension of the $V$ in $R^{2}$ to $R^{2}\times
Z_{2}$ is given by
\begin{equation}
\mathcal{V}=\left(
\begin{array}
[c]{cc}%
V_{L} & 0\\
0 & V_{R}%
\end{array}
\right)  ,\ V_{L}^{\dag}V_{L}=1,\ V_{R}^{\dag}V_{R}=1
\end{equation}
with appropriate matrices $V_{L}$ and $V_{R}$. Then the gauge connection
becomes%
\begin{equation}
\mathcal{A}=\mathcal{V}^{\dag}d\mathcal{V}=\left(
\begin{array}
[c]{cc}%
V_{L}^{\dag}dV_{L} & i\left(  V_{L}^{\dag}V_{R}-1\right) \\
i\left(  V_{R}^{\dag}V_{L}-1\right)  & V_{R}^{\dag}dV_{R}%
\end{array}
\right)  .
\end{equation}
We can see that the gauge connection discussed in this article is obtained
with the assignment
\begin{align}
L  &  =V_{L}^{\dag}dV_{L},\ R=V_{R}^{\dag}dV_{R},\nonumber\\
\varphi &  =V_{L}^{\dag}V_{R}-1,\nonumber\\
\phi &  =V_{L}^{\dag}V_{R}.
\end{align}
Therefore, we can complete the ADHM method, if we can introduce suitable
coordinate expression.

It is also natural to expect that the ADHM is applicable here on the ground
that there has been an analogy between the Yang equation for instanton and the
master equation for vortex. Let $z\equiv x_{1}+ix_{2},$ $w\equiv x_{3}+ix_{4}$
be complex coordinates in $R^{4},$ then the instanton equation, that is, the
anti-selfdual Yang-Mills equation is equivalent to the Yang equation%
\begin{equation}
\partial_{z}\left(  J^{-1}\bar{\partial}_{z}J\right)  +\partial_{w}\left(
J^{-1}\bar{\partial}_{w}J\right)  =0
\end{equation}
for the Yang's potential $J$ \cite{Yang:1977zf}. It is obvious that the master
equation is the analog of the Yang equation. The relations would be explained,
if we can regard $w$ as a coordinate of the $Z_{2}$ space and could assign
appropriate expression to them.

Instanton equation in the usual space can be completely solved by the ADHM
method. As for the vortex equation on the other hand, although it can be
rewritten as a master equation plus half-ADHM, the solution cannot be obtained
because we are left with the master equation. We consider however, that the
difference can be attributed to the structure of $Z_{2}$ space. And in order
to understand the situation, we have to have the representation of not only
the differential forms but the representation of the coordinates. For the
differential forms, we are using the matrix representation, but the
representation for the background noncommutative coordinates should be
considered separately. A construction method in terms of the coordinate will
appear in the work in preparation.

In this article, we have shown that the non-Abelian vortex in $R^{2}$ is
equivalent to the instanton on $R^{2}\times Z_{2}$ space. It has been proposed
in ref \cite{Popov}, that there exists similar relation in the case of the
model on compact Riemann surface $\Sigma$. They have shown that the instanton
on $\Sigma\times CP^{1}$ can be considered as a non-Abelian vortex on $\Sigma
$. It would be interesting to examine the relations between our work and their approach.

\bigskip\noindent{\Large \textbf{Acknowledgments}}

We would like to thank Akihiro Nakayama for his support and hospitality.

\providecommand{\href}[2]{#2}\begingroup\raggedright

\endgroup

\end{document}